\documentclass[twocolumn,showpacs,preprintnumbers,amsmath,amssymb,prb,superscriptaddress]{revtex4}

\usepackage{amsmath}
\usepackage{graphicx}
\usepackage{amssymb}
\usepackage{epstopdf}

\usepackage{color}
\usepackage{ulem}

\usepackage{verbatim} 

\begin{document}

\title{Microscopic theory of the insulating electronic ground states of actinide dioxides AnO$_2$  (An=U, Np, Pu, Am, and Cm)}

\author{M.-T.\ Suzuki}
\affiliation{CCSE, Japan Atomic Energy Agency, 5-1-5, Kashiwanoha, Kashiwa, Chiba 277-8587, Japan}
\affiliation{Department of Physics and Astronomy, Uppsala University, Box 516, SE-751 20 Uppsala, Sweden}
\author{N.\ Magnani}
\affiliation{Actinide Chemistry Group, Chemical Sciences Division, Lawrence Berkeley National Laboratory, 1 Cyclotron Road,
Berkeley CA 94720-8175, USA}
\author{P.\ M.\ Oppeneer}
\affiliation{Department of Physics and Astronomy, Uppsala University, Box 516, SE-751 20 Uppsala, Sweden}

\date{\today}
\begin{abstract}
 The electronic ground states of the actinide dioxides AnO$_2$ (with An=U, Np, Pu, Am, and Cm) are investigated employing first-principles calculations within the framework of the local density approximation +$U$ (LDA+$U$) approach, implemented in a full-potential linearized augmented plane-wave scheme.
A systematic analysis of the An-$5f$ states is performed which provides intuitive connections between the electronic structures and the local crystalline fields of the $f$ states in the AnO$_2$ series.
Particularly the mechanisms leading to the experimentally observed insulating ground states are investigated. These are found to be caused by the strong spin-orbit and Coulomb interactions of the $5f$ orbitals; however, as a result of the different configurations, this mechanism works in distinctly different ways for each of the AnO$_2$ compounds. 
In agreement with experimental observations, the nonmagnetic states of plutonium and curium dioxide are computed to be insulating, whereas those of uranium, neptunium, and americium dioxides require additional symmetry breaking to reproduce the insulator ground states, a condition which is met with magnetic phase transitions. We show that the occupancy of the An-$f$ orbitals is closely connected to each of the appearing insulating mechanisms. We furthermore investigate the detailed constitution of the noncollinear multipolar moments for transverse 3$\boldsymbol{q}$ magnetic ordered states in UO$_2$ and longitudinal 3$\boldsymbol{q}$ high-rank multipolar ordered states in NpO$_2$ and AmO$_2$.

\end{abstract}

\pacs{71.20.Nr, 71.27.+a, 75.30.Et}
\keywords{Actinide dioxides, multipole order, LDA+{\it U}}

\maketitle

\section{\label{sec:level1}INTRODUCTION}

Recently, $f$ electron materials containing actinide elements have drawn considerable interest, stimulated by observations of intriguingly ordered ground states forming at low temperatures.  A striking example is the formation of the hidden order phase in URu$_2$Si$_2$, where the origin of the arising low-temperature  electronic order could not unambiguously be disclosed even after a quarter century (see, e.g., Ref.\ \onlinecite{mydosh11}). High-rank multipoles have been proposed recently
as possible candidates for the order parameter in the hidden order phase.\cite{cricchio09,haule09,harima10,thalmeier11,kusunose11,ikeda12}
Another example is the ordered ground state of NpO$_2$, which, after many years could experimentally be established to be due to a high-rank multipolar order, in the absence of any dipolar moment formation.\cite{magnani08,santini09a}
The richness of $5f$ electron physics can be attributed to their multiple degrees of freedom where entangled spin and orbital moments may occur, activated through the strong spin-orbit (S-O) interaction in the actinide elements as well as on-site Coulomb interactions.
These conditions are in particular met in the actinide dioxides, which have provided a treasure trove of a rich variety of multi-orbital physics over many years.\cite{shiina97, kusunose08, kuramoto09, santini09a}

The actinide dioxides crystallize in the cubic fluorite structure, yet, the ground-state properties of the actinide dioxides are intriguingly diverse. The current knowledge of their ordered ground states stipulates that in UO$_2$ transverse 3$\boldsymbol{q}$  magnetic dipolar and 3$\boldsymbol{q}$  electric quadrupolar order is realized;\cite{amoretti89, carretta10} NpO$_2$ is characterized by a 3$\boldsymbol{q}$ ordered magnetic multipole of a high rank, in the absence of any dipole moment formation.
\cite{magnani08,santini09a}
AmO$_2$ undergoes a conspicuous phase transition at around 8.5~K.\cite{karraker75} While a peak structure in the magnetic susceptibility was found,\cite{karraker75} neutron-diffraction measurements could not detected antiferromagnetic order in agreement with the M{\"o}ssbauer measurement.\cite{kalvius69, boeuf79}
PuO$_2$ is a simple nonmagnetic insulator,\cite{yasuoka12} whereas CmO$_2$ has an unexpected paramagnetic moment.~\cite{morss89}
Surprisingly, each of these actinide dioxides is an insulator, in spite of the very different emerging orders and physical properties. This suggests that in these dioxides  different gap-formation mechanisms are in fact operative. The origin of gap formation can only restrictedly be established through experiments. Conversely, first-principles electronic structure calculations are well suited to study the gap formation mechanisms in relation to the unusually ordered states, but only a few such first-principles calculations, accounting for multipolar order states have been performed.\cite{suzuki10_1}

 A framework of first-principles calculations allows for a description of the bulk properties even if the investigated system contains some localized $f$ electrons in an open shell. However, it is known that the localized character of $f$ electrons is not reproduced with the basic approximations employed in first-principles calculations, like the local density approximation (LDA) or  generalized gradient approximation (GGA) in which the two-body correlation between electrons in the open shell is not captured sufficiently.
 It was actually reported that the LDA approach led to states that are inconsistent with the CEF ground states determined experimentally in the AnO$_2$.\cite{maehira07, hasegawa13} Moreover, the calculations predicted metallic ground states~\cite{petit96, maehira07} whereas the AnO$_2$ compounds with light actinides are known to be large gap semi-conductors or insulators with energy gaps around 2 eV.\cite{mcneilly64,schoenes80,McCleskey13}
Therefore,  first-principles calculations taking into account the strong Coulomb interaction have been applied  and have provided  insight in the detailed magnetic and electronic character of UO$_2$,~\cite{brooks83, kudin02, laskowski04, prodan07, yu11, zhou11, prodan06, Wen12} of NpO$_2$,~\cite{prodan07, suzuki10_1, Wen12} and of PuO$_2$,~\cite{colarieti02, petit03, prodan06, prodan07, jomard08, sun08, nakamura11, modin11, wang12, Wen12} although the $f$ characters provided in some of these calculations are different from those of the experimental ground states, especially for the ordered states.
As taking strong electron correlations into account is necessary to obtain the energy gap in the first-principles calculations, the insulating ground states of AnO$_2$ are referred to as  Mott insulators.~\cite{laskowski04, yin08, Yin11} 
 Thus, to appropriately describe the ground states in the AnO$_2$ compounds, it is imperative to include the effects of strong Coulomb interaction, strong spin-orbit coupling, and the multiorbital character of the $f$ electrons on equal footing.
 In this paper, we perform such first-principles calculations, focusing particularly on the mechanisms of insulating gap formation. We show that the insulating mechanisms differ substantially from one AnO$_2$ compound to another, depending on the involved $f$ states in AnO$_2$.

The LDA+$U$ method with S-O interaction included has been successful in capturing the ground state properties of $f$-electron compounds.~\cite{shick01_1, harima01, harima02, ghosh05, shorikov05, shick05, shick06, suzuki10_2, suzuki10_1} In spite of some limitations this method is especially suited to describe the local character of $f$ electrons on the same footing with the electronic band description.
 Moreover, there are two specific properties of the LDA+$U$ framework that make it suitable to study complex ordered phases in the AnO$_2$: the spin-orbital dependence of the local potential, which is essential since appearing order parameters involve multiple spin and orbital degrees of freedom, and the ability to take into account noncollinearity of local order parameters, because the multipole moments on each An site can be expressed through the local potential only.\cite{suzuki10_1}
Here  we have adopted the LDA+$U$ method with spin-orbit interaction, combined with the framework of the full-potential linearized augmented plane wave (FLAPW) band-structure method. Since a large Coulomb $U$ tends to increase the anisotropic character of the $f$ states, the full-potential treatment is an important ingredient, too, to adequately reproduce the  $f$ states in the AnO$_2$.

In the following, we  provide a complete computational study of the electronic structures of the AnO$_2$, focusing on the activated spin-orbital degree of freedom in the ground states. 
First, we give in Sec.\  \ref{exp-prop} an outline of the known experimental properties of the AnO$_2$ compounds.
In Sec. \ref{sec:level2}, we describe the employed computational framework, and in Sec. \ref{sec:level3}, we report the obtained results and provide discussions separately for our calculations of the nonordered states in the AnO$_2$ and for the ordered states computed for UO$_2$, NpO$_2$, and AmO$_2$.

\section{Properties of $\bf AnO_2$ compounds \label{exp-prop}}

\begin{figure}[tb]
 \begin{center}
  \includegraphics[width=0.8\linewidth]{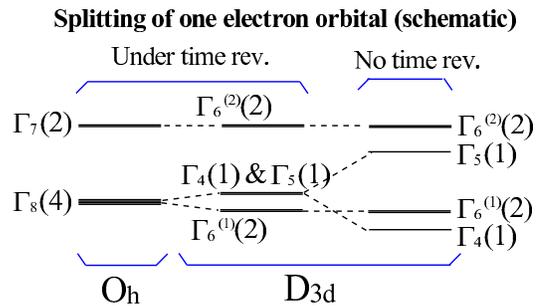}
  \caption{Schematic description of the actinide one-electron $j$=5/2 orbitals in the crystal fields with $O_h$ and $D_{3d}$ point-group symmetry, respectively.
  The  $\Gamma_8$ quartet in cubic $O_h$ symmetry splits in two singlets,  $\Gamma_4$ and $\Gamma_5$, degenerate under time reversal symmetry, and one doublet state, $\Gamma_{6}^{(1)}$, in $D_{3d}$ symmetry.}
\label{Fig:CEF}
 \end{center}
\end{figure}

The actinide dioxides crystallize in the cubic fluorite crystal structure. They are characterized by an ionicity with An$^{4+}$ cations and O$^{2-}$ anions, which is strong for the early $5f$ elements, but is reduced for the late actinide elements. \cite{prodan07} The dioxides with open $5f$ shell are further characterized by the strong localized character of the $5f$ electrons on the An atoms. Note that ThO$_2$ has no occupied $f$ orbitals and is therefore not considered here.

 
 The An-$5f$ states multiplets are classified as a $\Gamma_5$ triplet, a $\Gamma_8$ quartet, and a $\Gamma_1$ singlet in UO$_2$,\cite{rahman66, kern85, amoretti89, magnani05} NpO$_2$,\cite{fournier91, amoretti92, magnani05} and PuO$_2$,\cite{kern90, kern99, magnani05} respectively, based on the Russell-Saunders LS coupling scheme for their paramagnetic states.
 In a one electron description, the degenerate $5f$ orbitals split in $j$=$\frac{5}{2}$ and $j$=$\frac{7}{2}$ orbitals under the action of strong S-O coupling. The $j$=$\frac{5}{2}$ orbitals, which have a lower orbital energy than the $j$=$\frac{7}{2}$ orbitals, further split into a $\Gamma_7$ doublet and a $\Gamma_8$ quartet through the cubic crystalline electric field (CEF), see the schematic description in Fig.\ \ref{Fig:CEF}.
 Following a simple point charge model, the energy level of the $| j=5/2,\, \Gamma_8 \rangle$ quartet should be lower than that of the $| j=5/2, \, \Gamma_7 \rangle$ doublet due to the oxygen anions located in [1\,1\,1] directions from the actinide cations. 
If the CEF is large enough to neglect the higher-lying $\Gamma_7$ orbitals, the $f$ states can be described by states in which  two, three, and four electrons occupy the $\Gamma_8$ orbitals for U$^{4+}$, Np$^{4+}$, and Pu$^{4+}$ ions, respectively.
Although the simple one electron description qualitatively explains the local character of the $f$ states on the An single ion in these compounds,\cite{kubo05a, kubo05b} it is nonetheless necessary to analyze the electronic structure on the basis of first-principles calculations to reveal the bulk properties that involve the conduction as well as localized electrons, such as, e.g., the insulating behavior of the AnO$_2$.\cite{prodan06,prodan07,suzuki10_1}


 UO$_2$ exhibits a clear phase transitions at $T=30.8$ K.~\cite{osborne53, willis65}
The low temperature phase of UO$_2$ is associated with a transverse 3$\boldsymbol{q}$ antiferromagnetic structure,\cite{wilkins04, wilkins06, carretta10} which superseded the earlier suggestions of 1$\boldsymbol{k}$ (Ref.\ \onlinecite{allen68}) or 2$\boldsymbol{k}$ (Ref.\ \onlinecite{faber75}) structures,  with
a dipolar magnetic uranium moment of 1.74\,$\mu_{B}$.\cite{frazer65, faber76}
 The dipolar 3$\boldsymbol{q}$ magnetic order in UO$_2$ is accompanied by a 3$\boldsymbol{q}$ ordering of quadrupoles, which corresponds to a distortion of the $5f$ charge density along the direction of dipolar moment.\cite{wilkins06, carretta10}
 It has been proposed that the superexchange quadrupolar interaction is an important ingredient to stabilize the 3$\boldsymbol{q}$ magnetic ground state\cite{giannozzi87, mironov03, wilkins06} as well as for the magnetization dynamical properties observed for UO$_2$.\cite{carretta10, caciuffo11} Also, recent first-principles studies based on an LDA+$U$ method successfully showed the stability of the 3$\boldsymbol{q}$ dipolar magnetic ordered state in UO$_2$.\cite{laskowski04, zhou11}

The phase transition observed \cite{osborne53, ross67} at $T=25.4$ K in NpO$_2$ has been an enigmatic issue until not too long ago. At an earlier stage of the research on this issue, the low temperature phase of NpO$_2$ had been speculated to arise from an antiferromagnetic order due to a similar behavior of the temperature dependent susceptibility to that of UO$_2$. However, no dipole magnetic moments essential to characterize the phase transition have been observed in neutron scattering~\cite{fournier91, amoretti92} and M{\"o}ssbauer experiments.~\cite{dunlap68, friedt85, masaki00}
Intensive experimental and theoretical efforts then were invested to identify the unusual ordered state. Crucial experiments to identify aspects of the ordered state were the muon spin rotation measurement which detected breaking of the time reversal symmetry~\cite{kopmann98} and resonant x-ray scattering which identified the electric symmetry of the low temperature phase.\cite{paixao02} Several experiments are consistent with, and have confirmed, the noncollinear 3$\boldsymbol{q}$ multipolar order (MPO) with $\Gamma_{5}$ symmetry multipoles.\cite{fournier91, kopmann98, santini00, paixao02, santini02, caciuffo03, kiss03, lovesey03, sakai03, sakai05, nagao05, tokunaga05a, tokunaga05b, tokunaga06a, santini06}
Note that the $\Gamma_5$ multipoles do not contain a dipolar moment in its tensor elements.
In principle, the $\Gamma_{5}$ octupole\cite{note} moments could be detectable in resonant x-ray measurements,~\cite{lovesey05} however, direct observation of the octupole moments has not been successful so far. 
Several recent theories pointed out that a multipolar ordered state with triakontadipole (rank 5) primary order parameter would explain the small weight of the octupole moments.\cite{magnani08, santini09a, suzuki10_1}

 The phase transitions to the 3$\boldsymbol{q}$ MPO states break translation symmetry of the face centered cubic (FCC) crystal structure but preserve a simple cubic (SC) symmetry containing a four-sublattice unit cell.\cite{burlet86, ikushima01, paixao02}
A symmetry analysis of the low temperature phases observed in UO$_2$ and NpO$_2$ has already been performed.\cite{kiss03, nikolaev03, matteo07}
The space group $Fm\bar{3}m$ is lowered to the $Pa\bar{3}$ and the $Pn\bar{3}m$ groups in the ordered phases of UO$_2$ and NpO$_2$, respectively.\cite{paixao02,nikolaev03} 
 Since the magnetic space group of the 3$\boldsymbol{q}$-MPO states have no nonunitary part, both electric and magnetic multipoles which belong to the same symmetry can spontaneously appear in these ordered states.
The four-sublattice unit cell contains one site of equivalent eight oxygen atoms which has one Wycoff parameter in the $Pa\bar{3}$ and two nonequivalent fixed oxygen sites (cubic 2a site and tetragonal 6d site) in the $Pn\bar{3}m$ space group. These facts are clearly reflected in the measured $^{17}$O-NMR spectra.~\cite{sakai05, tokunaga06b} Although the antiferromagnetically ordered state in UO$_2$ would allow the oxygen atoms to move with the Wycoff parameter, it was reported that the deformation of the oxygen position is considerably small.\cite{faber76}

In PuO$_2$, the magnetic susceptibility is independent of temperature up to 1000~K.\cite{raphael68} The nonmagnetic property is understood from the CEF analysis for the Pu$^{4+}$ ion giving that the ground state is the nonmagnetic singlet and the first excited state, which is 123 meV above the ground state, is a triplet.\cite{kern90, kern99, santini99}
Transport measurements, furthermore, showed that PuO$_2$ is an insulator with a 1.8-eV activation gap,\cite{mcneilly64} while optical spectroscopy gave a 2.8-eV direct gap.\cite{McCleskey13}
A recent Pu-NMR study confirmed the nonmagnetic character of the ground state of PuO$_2$.~\cite{yasuoka12}

 In AmO$_2$ a clear antiferromagnetic-looking phase transition is observed at 8.5~K,\cite{karraker75} however, no magnetic moment has been observed in M{\"o}ssbauer nor in neutron scattering measurements.\cite{kalvius69, boeuf79}
The fingerprint of this dioxide is similar to the one obtained in earlier studies of NpO$_2$, and hence, an AF-MPO state is expected in AmO$_2$. 
The CEF ground states of the Am$^{4+}$ ion in AmO$_2$ have been believed to be a $\Gamma_7$ doublet.~\cite{karraker75, abraham71, kolbe74}
However, the $\Gamma_7$ state has no degree of freedom for the higher rank multipoles and therefore seems to contradict the experiments which observe no magnetic dipole moments.
A recent CEF analysis based on the $j$-$j$ coupling method discussed an instability of the $\Gamma_7$ ground state and possibility of stabilization of the $\Gamma_8$ ground state, which can induce higher order multipoles without inducing a dipole moment.\cite{hotta09} Notably, there are many existing experimental challenges to distinguish the essential bulk contribution of AmO$_2$ due to the strong self-radiation damage caused by alpha decay.\cite{benedict80, edelstein06, tokunaga10, tokunaga11}

For the next actinide dioxide, CmO$_2$, only a few experiments have thus far been reported for the detailed constitution of its ground state.\cite{morss89, kvashnina07}
Cm$^{4+}$ has six electrons in the 5$f$ shell, producing a nonmagnetic $^{7}$F$_{0}$ ground state from Hund's rules.
However, a paramagnetic moment has been observed in CmO$_2$ which is unexpected from the nonmagnetic ground state.~\cite{morss89} 
 Niikura and Hotta explained the magnetic behavior of CmO$_2$ by assuming the proximity of a magnetic excited state with an excitation energy smaller than the Land$\acute{\rm e}$ interval rule.\cite{niikura11} 

\section{\label{sec:level2} Computational Method}
\label{Sec:Method}

The LDA+$U$ method~\cite{anisimov91, czyzyk94, anisimov97, shick99} provides the one electron Hamiltonian as
\begin{eqnarray}
h_{{\rm LDA}+U}=h_{\rm LDA}+\sum_{\tau}\sum_{\gamma\gamma'}|\tau\ell\gamma\rangle v_{\gamma\gamma'}^{\tau\ell} \langle \tau\ell\gamma'|\ ,
\label{Eq:oneHamilton}
\end{eqnarray}
where the on-site $+U$ potential is given by
\begin{eqnarray}
\!\! \!\! v_{\{ms\}\{m's'\}}^{\tau\ell} &=&\sum_{m''m'''} \Big[ \delta_{ss'} \sum_{s''}n^{\tau\ell}_{\{m'''s''\}\{m''s''\}} \times \nonumber \\
& & \langle mm''|W|m'm'''\rangle \nonumber \\
&-&n_{\{m'''s\}\{m''s'\}}^{\tau\ell}\langle mm''|W|m'''m'\rangle \Big] \nonumber \\
&-&\delta_{mm'}\delta_{ss'} \big[ U(n-1/2)-\frac{J}{2}(n-1)\big] ,
\label{eq:Upotential}
\end{eqnarray}
where $\tau$ and $\ell$ denote the atoms and angular momenta of the orbitals, respectively, for which the $+U$ potentials are introduced. $\gamma$ ($\gamma'$) is an index related to an orbital $m$ ($m^{\prime}$) and a spin $s$ ($s'$), or, alternatively, double-valued irreducible representations of the site symmetry of the An site obtained through a unitary transformation. The matrix elements of the full Coulomb interaction of two $f$ electrons is expressed as
\begin{eqnarray}
\langle m_1m_2\mid W \mid m_3m_4\rangle =\delta_{m_1+m_2,m_3+m_4}\nonumber\\
\times\sum_{k=0}^{6}c^k (\ell m_1;\ell m_3)c^k(\ell m_4;\ell m_2)F^k\ ,
\end{eqnarray}
where the $F^k$ are the Slater-Condon parameters,~\cite{slater29,condon31} and $c_k$ is the Gaunt coefficient.~\cite{gaunt29,racah42}
In our calculations, $F_0$ is taken as $F_0=U$, and the Hund's coupling parameter $J$ is related with the higher order Slater integrals as $J=(286F_2+195F_4+250F_6)/6435$ for $f$ electrons.
The local potentials within the muffin-tin (MT) spheres are determined from the spin-orbital dependent density matrix, 
\begin{eqnarray}
n_{\gamma\gamma'}^{\tau\ell} = \int_{\rm MT} dr^{\tau} \{r^{\tau}\}^2 \rho^{\tau\ell}_{\gamma\gamma'}(r^{\tau}) ,
\end{eqnarray}
which contains a projection from the band states $| \boldsymbol{k}b \rangle$ to the local basis $|\tau\ell\gamma\rangle$,
\begin{eqnarray}
\rho^{\tau\ell}_{\gamma\gamma'}(r^{\tau})= \frac{1}{N} \sum_{\boldsymbol{k}b}\langle \tau\ell\gamma | \boldsymbol{k}b \rangle \ \langle \boldsymbol{k}b | \tau\ell\gamma'\rangle ,
\end{eqnarray}
$N$ being the number of $\boldsymbol{k}$ points and $r^{\tau}$ the radial component of the position vector
$\boldsymbol{r}^{\tau}$ measured from atom $\tau$. The density matrix as well as the charge density are determined selfconsistently.

The calculations have been performed for nonordered states in the series of AnO$_2$ and for ordered states in UO$_2$, NpO$_2$, and AmO$_2$.
 In the calculations of nonordered states, we used the FCC unit cell with the space group $Fm\bar{3}m$ (No.\ 225) and applied a relation $n_{-m-m'-s-s'}^{i\ell}$=$(-1)^{m+m'+s-s'}$$n_{mm'ss'}^{i\ell*}$, imposed  by time-reversal symmetry.~\cite{harima01, suzuki10_2}
 The ordered states are calculated with the four sub-lattice unit cell mentioned above.
 It is known that the large $U$ introduced in the LDA+$U$ method can induce some meta-stable states especially in calculations of ordered states and may lead to convergence to an electronic state that is inconsistent with the realistic ground state.~\cite{shick01_2, larson07, jomard08, dorado09, dorado11}
 To avoid this problem, we made use of experimental information concerning the CEF ground states and the order parameters to control the occupations for the initial density matrix and adapt the symmetry preserving the 3$\boldsymbol{q}$ structure of the order parameters observed for UO$_2$ and NpO$_2$.
Doing so, we could confirm that the calculations successfully stabilized the ordered states after convergence of the charge density and the density matrices on the An sites.

The (noncollinear) magnetic multipole moments can be described through the local LDA+$U$ potentials. 
The expectation values of the local operators $O^{\tau \ell}$ defined on a spin-orbital space of the An atoms are calculated with the local basis set \{$|\tau\ell\gamma \rangle$\} inside the MT spheres, following the expressions
\begin{eqnarray}
O^{\tau\ell}( \boldsymbol{r}^{\tau}) &\equiv& \frac{1}{N^2}\sum_{\boldsymbol{k} b}\sum_{\boldsymbol{k}' b'}\sum_{\gamma\gamma'} \langle \boldsymbol{r}^{\tau} | \boldsymbol{k} b \rangle \ \langle \boldsymbol{k} b | \tau\ell\gamma\rangle O^{\tau\ell}_{\gamma\gamma'} \nonumber \\
&&\times \langle \tau\ell \gamma' | \boldsymbol{k}' b' \rangle \ \langle \boldsymbol{k}' b' | \boldsymbol{r}^{\tau}\rangle ,  
\label{Eq:MPDist}
\end{eqnarray}
and
\begin{eqnarray}
\!\!\! \!\! \langle O^{\tau\ell}\rangle &=& \!\!\!
\int_{\rm MT} d \boldsymbol{r}^{\tau} O^{\tau\ell}( \boldsymbol{r}^{\tau})  \nonumber \\
&=& \!\!\!\sum_{\gamma} \sum_{\gamma_1\gamma_2} \int \! dr^{\tau} \{r^{\tau}\}^{2}\rho^{\tau\ell}_{\gamma\gamma_1}(r^{\tau})O^{\tau\ell}_{\gamma_1\gamma_2}\rho^{\tau\ell}_{\gamma_2\gamma}(r^{\tau}) ,
\label{Eq:MPExpect}
\end{eqnarray}
within the space limited to an orbital $\ell$.
Explicit expressions for the local multipole operators have been listed by Kusunose.~\cite{kusunose08}
We used the exchange-correlation functional of Gunnarsson and Lundqvist for the LDA potential.~\cite{gunnarsson76}
The Coulomb $U$ parameter has been chosen as $U=4$ eV and the exchange $J$ in the range of $0  - 0.5$ eV. 
These values have previously been shown to provide an accurate description of measured properties of actinide dioxides.\cite{dudarev97, dudarev98, modin11}
The double-counting term has been chosen as in the fully localized limit,\cite{liechtenstein95} leaving out the spin dependency of the Hund's coupling part to adapt it for the nonmagnetic LDA part of Eq.\ (\ref{Eq:oneHamilton}).

In the basis set we have included the Np 5$f$, 6$d$, 7$s$, and 6$p$ orbitals as valence states and Np 5$d$, 6$s$ orbitals as semi-core states and for oxygen we treated the 2$s$ and 2$p$ states as valence states. The MT sphere radii were 1.4 {\AA} for An and 0.9 {\AA} for O.
The  plane-wave cut-offs used in the calculations were about 250 and 900 plane waves at the $\Gamma$ point for calculations with the FCC normal cell  and for the 4-sublattice SC unit cell in the multipolar calculations, respectively. 
In reciprocal space we used for the self-consistent convergence (charge density and density of states (DOS) calculation) 12$\times$12$\times$12 (24$\times$24$\times$24) $\boldsymbol{k}$-points  in the FCC unit cell and 6$\times$6$\times$6 (12$\times$12$\times$12) for the 4-sublattice SC unit cell.

\section{\label{sec:level3} Results and discussion}

\subsection{Nonordered state calculations}

\begin{figure}[tb]
 \begin{center}
  \includegraphics[width=0.85\linewidth]{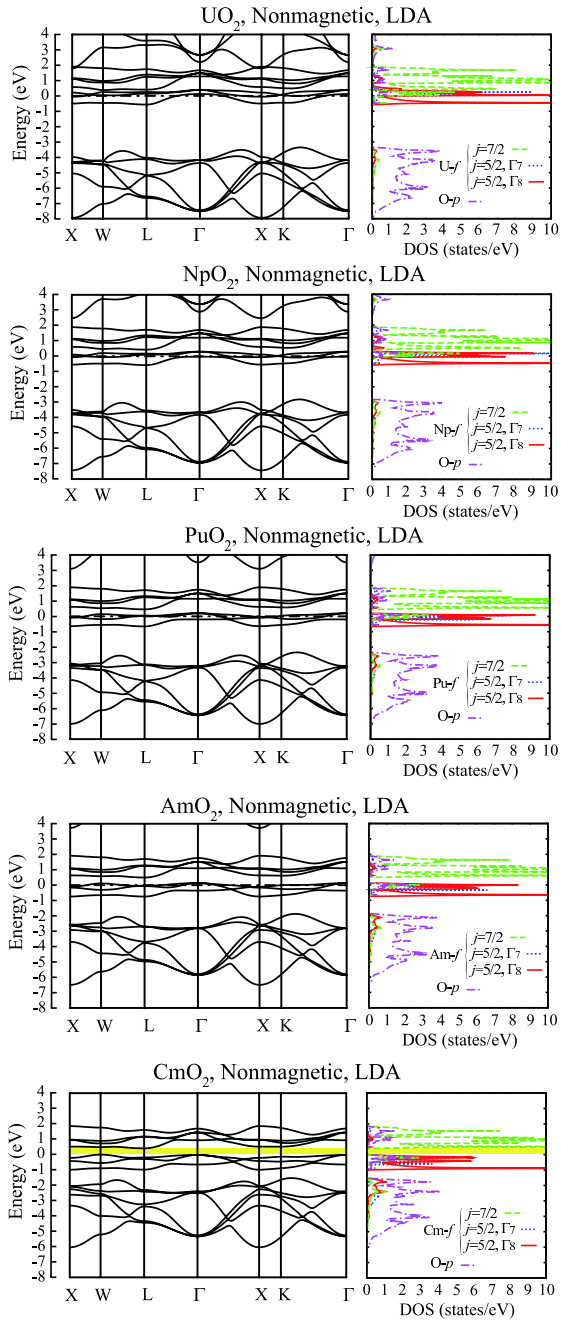}
  \caption{(Color online) Calculated band structures and density of states (DOS) for the nonmagnetic solution (see text) of the AnO$_2$, calculated by the LDA. The presence of a gap at the Fermi energy is indicated by the shaded area.
 The position of the Fermi energy is at 0 eV.}
\label{Fig:Band_NonOrder_LDA}
 \end{center}
\end{figure}

\begin{figure}[tb]
 \begin{center}
  \includegraphics[width=0.85\linewidth]{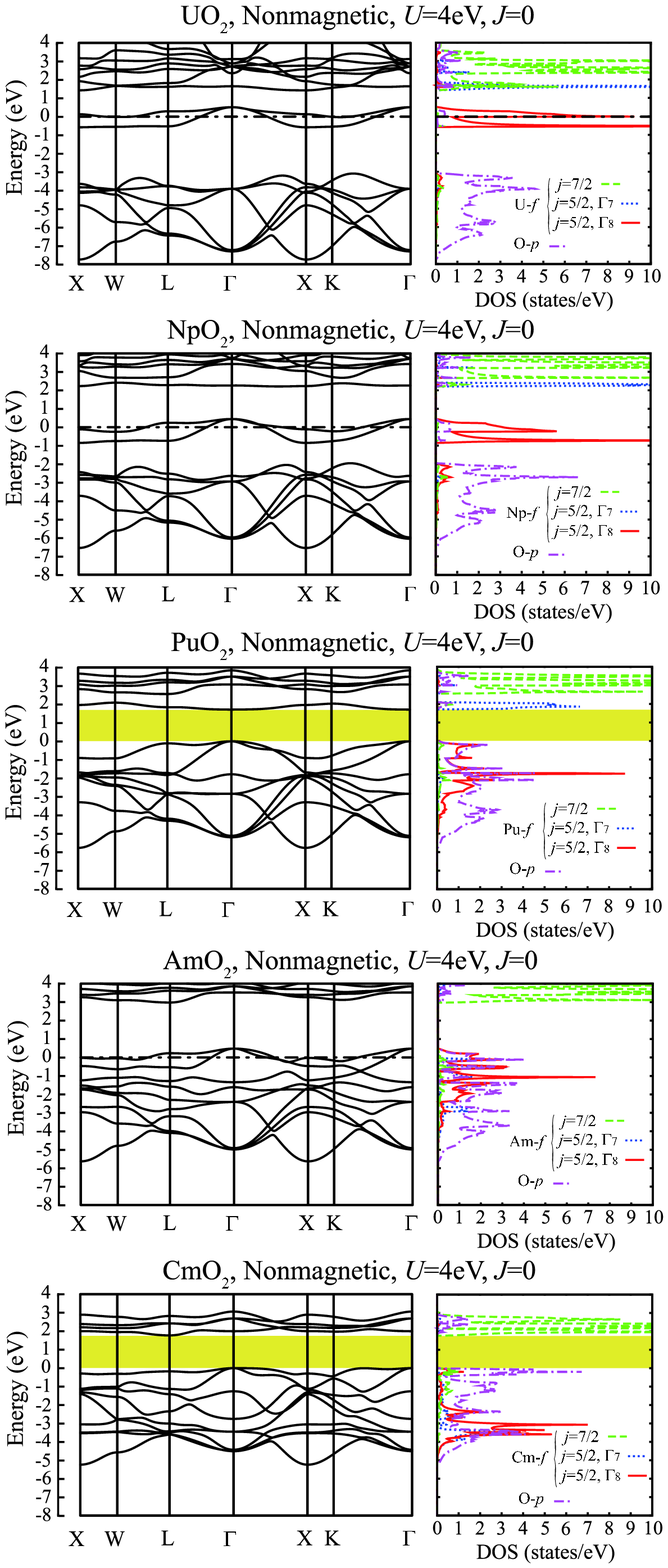}
  \caption{(Color online) As Fig.\ \ref{Fig:Band_NonOrder_LDA}, but calculated with the LDA+$U$ method ($U=4$ eV, $J=0$).}
\label{Fig:Band_NonOrder_U4J0}
 \end{center}
\end{figure}

\begin{figure}[tb]
 \begin{center}
  \includegraphics[width=0.85\linewidth]{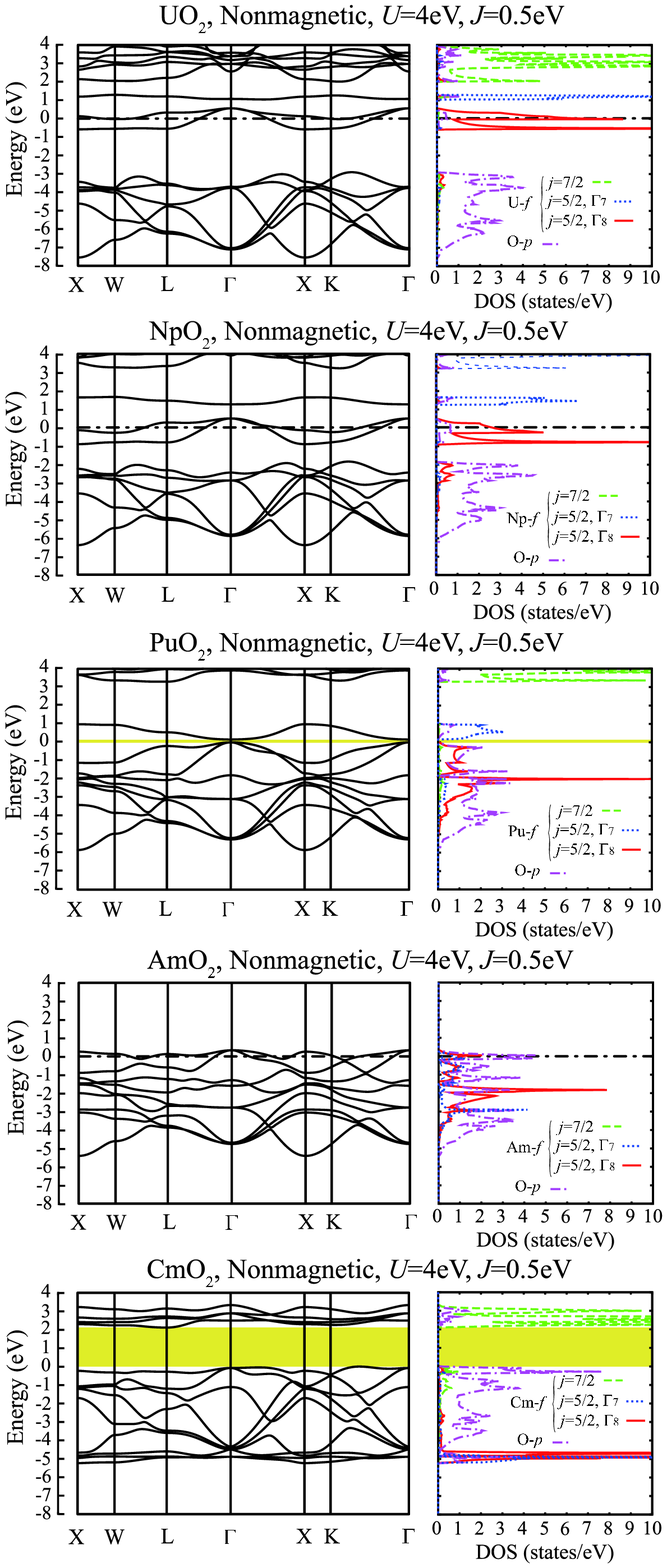}
  \caption{(Color online) As Fig.\ \ref{Fig:Band_NonOrder_LDA}, but calculated with the LDA+$U$ method ($U=4$ eV, $J=0.5$ eV).}
\label{Fig:Band_NonOrder_U4J05}
 \end{center}
\end{figure}

\begin{figure}[h]
 \begin{center}
 \includegraphics[width=0.55\linewidth]{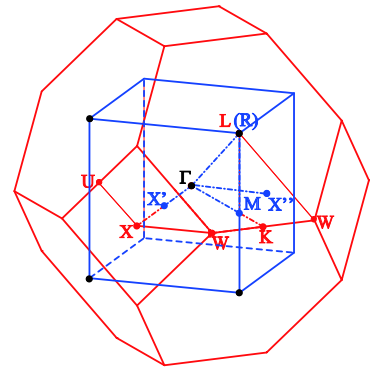}
  \caption{(Color online) The first Brillouin zones of the FCC unit cell used for calculations of nonmagnetic or nonmultipolar ordered actinide dioxides  (red lines) and the SC Brillouin zone of the 4-sublattice unit cell for 3$\boldsymbol{q}$-ordered states (blue lines). High-symmetry points are indicated.}
\label{Fig:brzone}
 \end{center}
\end{figure}

We first discuss the results obtained from the nonordered state calculations to examine how the anisotropic $f$ character is reproduced with the LDA+$U$ method.

In Figures \ref{Fig:Band_NonOrder_LDA}, \ref{Fig:Band_NonOrder_U4J0}, and \ref{Fig:Band_NonOrder_U4J05} we show the band structures and the DOS, respectively, calculated for the AnO$_2$ compound by the LDA, and by the LDA+$U$ method, with $U=4$ eV and $J=0$ and $J=0.5$ eV, respectively.
For comprehension we show in Fig.\ \ref{Fig:brzone}  the FCC BZ of the nonordered state, with high-symmetry points indicated.
Figure \ref{Fig:Band_NonOrder_LDA} exemplifies that the LDA approach only correctly predicts that nonordered CmO$_2$ would be an insulator. 
 In the LDA calculations, the $| j=\frac{5}{2}, \Gamma_7 \rangle$ and the $| j=\frac{5}{2}, \Gamma_8 \rangle$ states are present in the same energy range around the Fermi level, leading to a more or less homogeneous $f$ electron occupation for the $j =5/2$ orbitals. The computed orbital occupations of the $f$ states are given in Table \ref{tab:Occupation}. 
 As a consequence of the homogeneous occupation, the $f$ states predicted by the LDA calculations do not have sufficient anisotropic character as would be expected from the CEF states.  In other words, the LDA calculations fail to produce the CEF ground states in the AnO$_2$ series as was also previously pointed out,~\cite{maehira07, hasegawa13}
  whereas the correct CEF behavior for lighter actinide dioxides is reproduced if only the lower $\Gamma_8$ states are taken into account.

  \begin{center}
  \begin{ruledtabular}
  \begin{table*}[tbh]
      \caption{Calculated $5f$ electron occupation numbers per one-electron orbital on the An  sites in the AnO$_2$ (An=U, Np, Pu, Am, and Cm) fluorite-structure compounds.
      The calculations have been performed assuming the nonmagnetically ordered state, both with the LDA and with the LDA+$U$ ($U=4$ eV, and $J=0$ and 0.5 eV) approaches.
      }
    \label{tab:Occupation}

    \begin{tabular}{l l l c c c c c c} 
 \multicolumn{2}{l}{ }       & \multicolumn{3}{c}{$j$=5/2} & \multicolumn{4}{c}{$j$=7/2} \\ \cline{3-5} \cline{6-9}
 \multicolumn{1}{l}{} & Approach & $\Gamma_7$(2) & $\Gamma_8$(4)  & Total & $\Gamma_6$(2) & $\Gamma_7$(2) & $\Gamma_8$(4) & Total \tabularnewline[\doublerulesep]   \hline 
 UO$_2$   &   LDA          & 0.41 & 1.59 & 2.00 & 0.11 & 0.20 & 0.23 & 0.54 \\ 
          &  LDA+$U$, $J$=0 & 0.14 & 2.00 & 2.14 & 0.06 & 0.13 & 0.11 & 0.31 \\ 
          &  LDA+$U$, $J$=0.5\,eV & 0.18 & 2.02 & 2.21 & 0.06 & 0.13 & 0.10 & 0.29 \tabularnewline[\doublerulesep] \hline 
 NpO$_2$  &   LDA          & 0.81 & 2.18 & 3.00 & 0.11 & 0.26 & 0.27 & 0.65 \\ 
          &  LDA+$U$, $J$=0 & 0.16 & 3.09 & 3.25 & 0.06 & 0.06 & 0.29 & 0.41 \\ 
          &  LDA+$U$, $J$=0.5\,eV & 0.23  & 3.12 & 3.35 & 0.06 & 0.13 & 0.10 & 0.28 \tabularnewline[\doublerulesep] \hline 
 PuO$_2$  &   LDA          & 1.16 & 2.81 & 3.97 & 0.12 & 0.31 & 0.35 & 0.78 \\ 
          &  LDA+$U$, $J$=0 & 0.22 & 3.86 & 4.09 & 0.08 & 0.19 & 0.20 & 0.47 \\ 
          &  LDA+$U$, $J$=0.5\,eV & 0.41 & 3.87 & 4.29 & 0.07 & 0.14 & 0.13 & 0.34 \tabularnewline[\doublerulesep] \hline 
 AmO$_2$  &   LDA          & 1.54 & 3.34 & 4.88 & 0.15 & 0.37 & 0.44 & 0.96 \\ 
          &  LDA+$U$, $J$=0 & 1.91 & 3.44 & 5.34 & 0.07 & 0.26 & 0.13 & 0.46 \\ 
          &  LDA+$U$, $J$=0.5\,eV & 1.86 & 3.72 & 5.58 & 0.06 & 0.13 & 0.09 & 0.28 \tabularnewline[\doublerulesep] \hline 
 CmO$_2$  &   LDA          & 1.85 & 3.80 & 5.65 & 0.19 & 0.46 & 0.60 & 1.25 \\ 
          &  LDA+$U$, $J$=0 & 1.92 & 3.93 & 5.85 & 0.13 & 0.41 & 0.29 & 0.83 \\ 
          &  LDA+$U$, $J$=0.5\,eV & 1.97 & 3.94 & 5.92 & 0.11 & 0.30 & 0.25 & 0.66 \tabularnewline[\doublerulesep]
    \end{tabular}

  \end{table*}
  \end{ruledtabular}
  \end{center}

\begin{figure}[h]
 \begin{center}
 \includegraphics[width=1.0\linewidth]{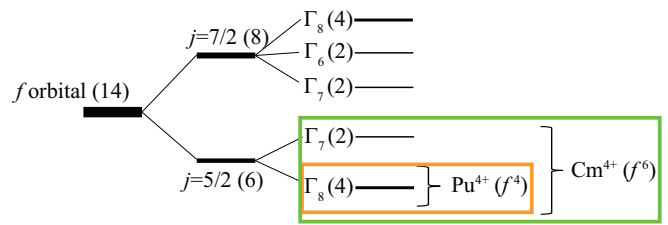}
  \caption{(Color online) Schematic description of spin-orbit split, nonmagnetic crystal-field ground states in terms of one-electron $f$ orbitals. Numbers in brackets give the degeneracy of the one-electron orbitals.}
\label{Fig:nmgGS}
 \end{center} 
\end{figure}

\begin{figure}[h]
 \begin{center}
\includegraphics[width=0.67\linewidth]{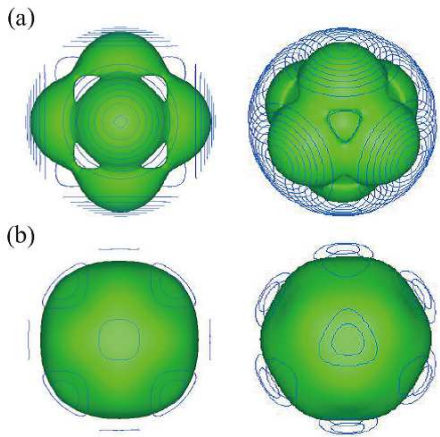}
  \caption{(Color online) Charge distributions of the An-5$f$ electrons (a) in PuO$_2$ and (b) in CmO$_2$ computed with the LDA+$U$  (using $U=4$ eV and $J=0.5$ eV). The distributions in the left-hand panels are viewed from the [1\,0\,0] direction, those in the right-hand panels from the [1\,1\,1] direction. The lines on the outside of the spheres represent the contour lines for the charge distribution (cf.\ Ref.\ \onlinecite{fazekas05}).}
\label{Fig:ChgSpn_PuCm}
 \end{center}
\end{figure}

Conversely, the LDA+$U$ calculations (Figs.\ \ref{Fig:Band_NonOrder_U4J0} and \ref{Fig:Band_NonOrder_U4J05}) do lead to anisotropic $f$ states, in which the $| j=5/2, \, \Gamma_8 \rangle$ quartet is dominantly occupied for An=U, Np, and Pu in AnO$_2$ due to a clear splitting between the $| j=5/2,\,\Gamma_7 \rangle$ and $| j =5/2,\,\Gamma_8 \rangle$ orbitals.

For PuO$_2$, an insulator state is obtained with the  $ |j=5/2,\, \Gamma_8\rangle$ orbitals being fully occupied, see Table \ref{tab:Occupation}.
 This state is consistent with the experimentally observed nonmagnetic insulator ground state with a singlet $\Gamma_1$ CEF state,~\cite{mcneilly64,yasuoka12} as is illustrated in the one-electron CEF states shown in Fig.\ \ref{Fig:nmgGS}.
Meanwhile, a previous first-principles study of PuO$_2$ reported that magnetic states are energetically more favorable than the nonmagnetic state for PuO$_2$ within LDA+$U$ calculations.~\cite{nakamura11} This is probably related to the approximations inherent to the LDA+$U$ method, in which the local exchange interaction for the electron Coulomb interaction is introduced within the Hartree-Fock approximation. The physical property of the experimentally observed nonmagnetic ground state should be captured with the computed nonmagnetic solutions, which reproduce the experimentally observed $\Gamma_1$ singlet crystal field ground state. In nonmagnetic LDA+$U$ calculations for PuO$_2$, the energy gaps are reduced with increasing $J$ as illustrated in Fig.\ \ref{Fig:Band_NonOrder_U4J05}. In this regard the $J$, which introduces Hund coupling-like electron interaction, appears to make the nonmagnetic insulator ground state less stable in PuO$_2$. In addition we mention that it has previously been shown that obtaining the nonmagnetic ground state of $\delta$-Pu with LDA+$U$ calculations depends sensitively on the type of the employed $+U$ functional.\cite{shick05}
The anisotropic character of the $f$ states in PuO$_2$ can be visually recognized from the $f$-charge distribution on the Pu sites shown in Fig.\ \ref{Fig:ChgSpn_PuCm}(a). It can be seen that the charge distribution is extended to the [1\,0\,0] and the equivalent axes, reflecting an $f$ wave function with $\Gamma_{8}$ symmetry.

 In UO$_2$ and in NpO$_2$ the strong ionization constrains the $f$ states to containing 2 or 3 electrons by the strong CEF effect. As a result, the $\Gamma_8$ orbitals being occupied with 2 or 3 electrons lead to metallic electronic states in UO$_2$ and in NpO$_2$ in the nonordered-state calculations. 
 
In AmO$_2$, the $\Gamma_7$ doublet is populated with about two electrons and the $\Gamma_8$ quartet has some hole (see Table \ref{tab:Occupation}), rendering also AmO$_2$ metallic in nonordered-state LDA+$U$ calculations. 
Note that in the LDA+$U$ calculations with $J=0.5$ eV the majority of empty Am $j=7/2$ states is located above 4 eV in Fig.\ \ref{Fig:Band_NonOrder_U4J05}.
Our results obtained from nonmagnetic calculations imply that the nonordered states in AmO$_2$ should exhibit a strong susceptibility due to the $\Gamma_8$ orbitals. 
The finding of an incompletely filled  $\Gamma_8$ orbital is consistent with the recent CEF theory suggesting that multipole order is realized within the $\Gamma_8$ CEF ground state in AmO$_2$.\cite{hotta09} 
Since the $\Gamma_{7}$ CEF ground states, which were suggested on the basis of experimental results,\cite{abraham71, kolbe74, karraker75} could not bring about any higher order multipoles, our results support the $\Gamma_8$ CEF ground states in AmO$_2$.
Furthermore, our results purport that symmetry breaking of some sort is necessary to reproduce the experimentally observed insulator ground states in UO$_2$, NpO$_2$, and AmO$_2$.
Especially, the insulator ground states can evidently not be produced with the nonordered-state calculations for NpO$_2$ and for AmO$_2$, since the FCC primitive cells containing an odd number of electrons will always lead to {\it uncompensated} metallic states when time-reversal symmetry is present.

\begin{figure}[tb]
 \begin{center}
  \includegraphics[width=0.88\linewidth]{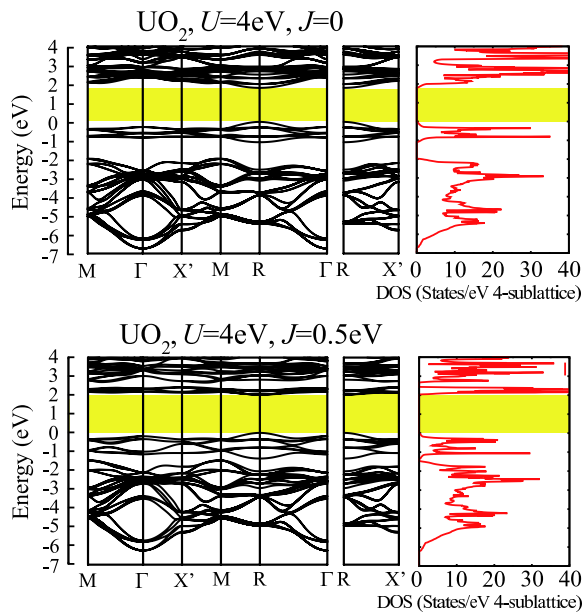}
  \caption{(Color online) Computed band structures and density of states in the transverse 3$\boldsymbol{q}$ $\Gamma_{4u}$ magnetic MPO state of UO$_2$, calculated with the LDA+$U$ method ($U=4$ eV, $J=0$ (top) and $J=0.5$ eV (bottom)). The gap computed at the Fermi energy is depicted by the shaded area.}
\label{Fig:BandDOS_UO2_Order}
 \end{center}
\end{figure}

\begin{figure}[htb]
 \begin{center}
  \includegraphics[width=1.0\linewidth]{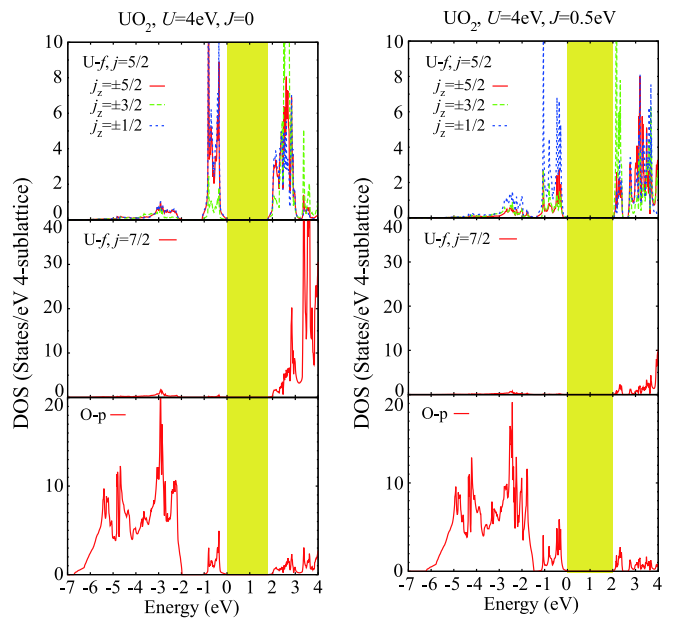}
  \caption{(Color online) The orbital-projected DOS components computed for the multipolar ordered phase of UO$_2$, for $J=0$ (left) and $J=0.5$ eV (right).}
\label{Fig:DOS_UO2_Order}
 \end{center}
\end{figure}

For CmO$_2$, both the LDA and the LDA+$U$ calculations produce the insulator states in the nonmagnetic-state calculations. 
We can speculate that a reason that the LDA approach is already a good approximation for CmO$_2$ comes from, first, the strong S-O splitting filling the $j=5/2$ manifold, and second, the weakly anisotropic character of the $f$ states in the nonmagnetic ground state, which can be visually seen in Fig.\ \ref{Fig:ChgSpn_PuCm}.
The Cm 5$f$ charge distribution, seen from the [1\,0\,0] direction [Fig.\ \ref{Fig:ChgSpn_PuCm}(b)], is much more isotropic than that of Pu [Fig.\  \ref{Fig:ChgSpn_PuCm}(a)].
Assuming that CmO$_2$ has a tetravalent Cm$^{4+}$ ion just as the An ion in the other actinide dioxides, a singlet ground state with $J=0$ is expected. In this regard, the nonmagnetic insulator states obtained in both calculations seem to be a natural consequence of the tetravalent ionized state in CmO$_2$. We note that energetically more favorable magnetic states in CmO$_2$ could be present due to the limitation of LDA+$U$ method just as in the case of PuO$_2$, but such magnetic states would be inconsistent with the expected nonmagnetic $\Gamma_1$ singlet ground state.
 A neutron diffraction experiment reported detection of an effective paramagnetic moment $\mu_{\rm eff} \sim 3.4$ $ \mu_{\rm B}$, which would be consistent with the Curie-Weiss behavior observed for the magnetic susceptibility.~\cite{morss89} Niikura and Hotta investigated the possibility of having a magnetic excited state just above the nonmagnetic ground state in the Cm$^{4+}$ ion with a small excitation energy, aiming to provide an understanding for the unexpected magnetic behavior in CmO$_2$.\cite{niikura11}

\subsection{Ordered state calculations}

Next, we have performed electronic structure calculations for the UO$_2$, NpO$_2$, and AmO$_2$ compounds allowing for self-consistent convergence to a symmetry-broken ordered ground state. 
 Figure \ref{Fig:brzone} shows the SC Brillouin zone (BZ) for the 3$\boldsymbol{q}$ ordered state, which displays a relationship to the FCC BZ of the nonordered state.
In the following the results of the ordered-state calculations are discussed in detail for each of these actinide dioxides.


\begin{figure}[tbh]
 \begin{center}
  \includegraphics[width=0.75\linewidth]{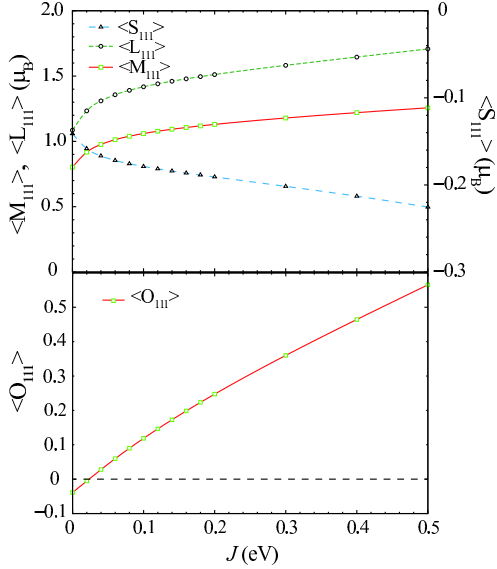}
  \caption{(Color online) Top: computed magnetic dipole moments in UO$_2$, as a function of the Hund's exchange $J$ parameter. Shown are the spin moment $\langle S_{111} \rangle$, the orbital moment $\langle L_{111} \rangle$, and the total moment, $\langle M_{111}\rangle=\langle L_{111}\rangle+2\langle S_{111}\rangle$ (see text).
  Bottom: the computed quadrupole moments $\langle O_{111}\rangle$=$\frac{1}{\sqrt{3}}[\langle O_{yz}\rangle+\langle O_{zx}\rangle+\langle O_{xy}\rangle ]$ in UO$_2$ as a function of $J$ applied to the $f$ electrons. }
\label{Fig:MagnQuad_UO2}
 \end{center}
\end{figure}

\begin{figure}[tbh]
 \begin{center}
 \includegraphics[width=0.7\linewidth]{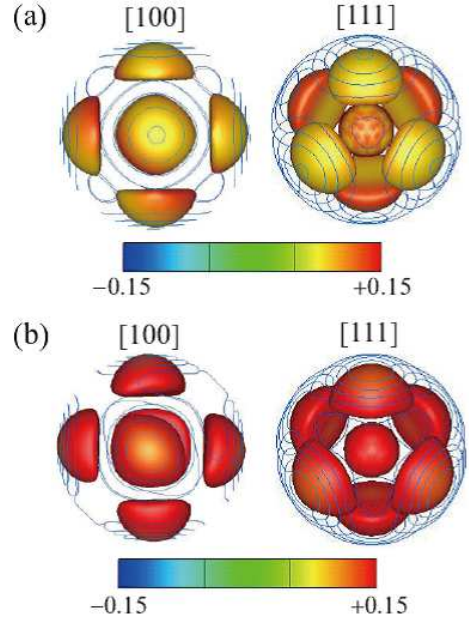}
  \caption{(Color online) Charge and magnetic distributions of the U-5$f$ electrons in UO$_2$, computed with (a) $U=4$ eV and $J=0$  and (b) $U=4$ eV and $J=0.5$ eV. The distributions in the left-hand panels are viewed from the [1\,0\,0] direction, and those in the right-hand panels from the [1\,1\,1] direction, corresponding to the threefold axis. The charge distributions are depicted by the isodensity surface, and the distributions of magnetic moments $M_{[111]}( \boldsymbol{r})=\frac{1}{\sqrt{3}} \{M_x( \boldsymbol{r})+M_y( \boldsymbol{r})+M_z(\boldsymbol{r})\}$ (shown by the color code) are calculated from Eq.\ (\ref{Eq:MPDist}) and plotted on the isosurface of the charge density. 
}
\label{Fig:ChgSpn_U}
 \end{center}
\end{figure}

 For UO$_2$, we choose the initial density matrix to correspond to the nonordered state with the Zeeman-type field along [1\,1\,1] equivalent directions, keeping a transverse-3$\boldsymbol{q}$ structure as symmetry breaking term.
The converged electronic structure corresponds to the transverse 3$\boldsymbol{q}$ ordered state with $\Gamma_4$ local multipoles.
The obtained band structures and DOS are shown in Figs.\ \ref{Fig:BandDOS_UO2_Order} and  \ref{Fig:DOS_UO2_Order}, for $U=4$ eV and two $J$ values, 0 and 0.5 eV, respectively.
 The energy gaps upon convergence are generated through the splitting of the $f$ states, which is in turn induced through the large $U$. In nonmagnetic UO$_2$ the plus and minus $j_z$ components of the $j_z$=$\pm$5/2, $\pm$3/2, $\pm$1/2 orbitals are degenerate; because of the Coulomb $U$ these orbitals split into the lower and upper Hubbard bands in the ordered states as seen in Fig.\ \ref{Fig:DOS_UO2_Order}. 
For $U=4$ eV the experimentally observed energy gap \cite{schoenes80} of about 2 eV is well reproduced with the 3$\boldsymbol{q}$ magnetic order in the calculations.
 
%

In the transverse 3$\boldsymbol{q}$ magnetic order, the dipolar magnetic moments are induced along [1\,1\,1] axis for the U atom at $(0,0,0)$, along the 
[-1\,1\,-1] axis for the U at $(0, 1/2, 1/2)$, along the [-1\,-1\,1] axis for the U at $(1/2, 0, 1/2)$, and along [1\,-1\,-1] for the U at $(1/2,1/2,0)$.
 The local magnetic moments on the uranium sites are shown in Fig.\ \ref{Fig:MagnQuad_UO2} as a function of the Hund's coupling parameter $J$. As seen from the figure, the magnetic moment is dominated by the orbital moment and is gradually enhanced for increasing $J$.
  The calculated orbital and spin moments are $\langle L_{111}\rangle = \frac{1}{\sqrt{3}}$ $ [ | \langle L_{x}\rangle | + |\langle L_{y}\rangle | + |\langle L_{z}\rangle | ]=1.09$ $\mu_B$ and $\langle S_{111}\rangle=\frac{1}{\sqrt{3}}$ $[ |\langle S_{x}\rangle | + | \langle S_{y}\rangle | + |\langle S_{z}\rangle |  ] =-0.14$ $\mu_B$ for $J =0$ and $\langle L_{111}\rangle =1.71$ and $\langle S_{111}\rangle =-0.23$ $\mu_B$ for $J =0.5$ eV. The total local magnetic moments are $\langle M\rangle = \langle L_{111}\rangle + 2 \langle S_{111} \rangle=0.80$ $ \mu_B$ for $J=0$ and 1.26 $\mu_B$ for $J=0.5$ eV, providing slightly smaller values compared with the experimental value \cite{faber76, schoenes80} of 1.74 $\mu_B$. Our results show that the complex constitution of the $f$ ground state, with contributions from the hybridization between U-5$f$ and O-2$p$ states as seen in Fig.\ \ref{Fig:DOS_UO2_Order}, leads to a reduction of the magnetic moment expected for the $\Gamma_5$ CEF ground state, which is slightly higher than 2.0 $\mu_B$.\cite{giannozzi87}
  We further provide in Fig.\ \ref{Fig:MagnQuad_UO2} the $J$ dependence of the quadrupole moments $\langle O_{111}\rangle = \frac{1}{\sqrt{3}}[\langle O_{yz}\rangle +\langle O_{zx}\rangle +\langle O_{xy}\rangle ]$, calculated from Eq.\ (\ref{Eq:MPExpect}) for the uranium $f$ electrons.
We find that the quadrupole moments also develop with increasing $J$, changing the sign around $J=0.03$ eV. The finite contribution from the quadrupole moments is consistent with recent experiments.~\cite{wilkins06, carretta10}

In Fig.\ \ref{Fig:ChgSpn_U} the spatial shape of the uranium $5f$-wave function is displayed by plotting its magnetic moment distribution projected to the [1\,1\,1] local axis on an isodensity surface of the calculated $f$-charge density for $U=4$ eV and two $J$ values, 0 and 0.5 eV. It is seen that an overall, dipolar magnetic moment exists along the [1\,1\,1] axis and that the local moments are enhanced with the effect of $J$. Also, reflecting the local site symmetry of the ordered state, the calculated charge and magnetic distributions preserve the $C_{3i}$ symmetry for UO$_2$.

\begin{figure}[tbh]
 \begin{center}
  \includegraphics[width=0.88\linewidth]{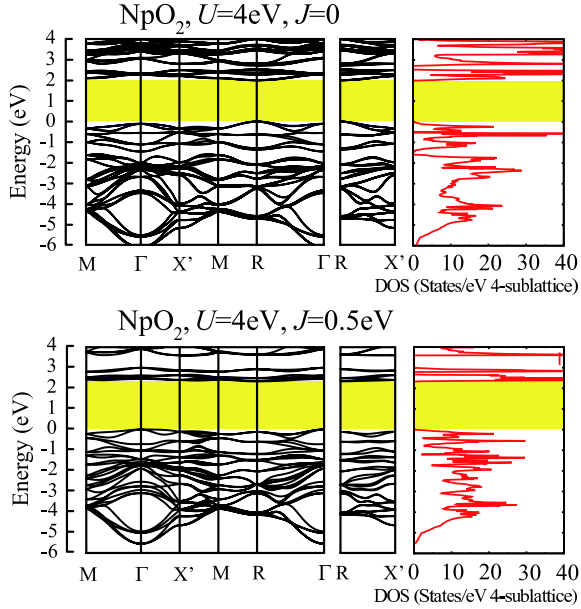}
  \caption{(Color online) Computed band structures and density of states in the longitudinal 3$\boldsymbol{q}$ $\Gamma_{5u}$ magnetic MPO state of NpO$_2$, calculated by the LDA+$U$ method ($U=4$ eV, $J=0$ or $J=0.5$ eV). }
\label{Fig:BandDOS_NpO2_Order}
 \end{center}
\end{figure}

\begin{figure}[tbh]
 \begin{center}
  \includegraphics[width=1.0\linewidth]{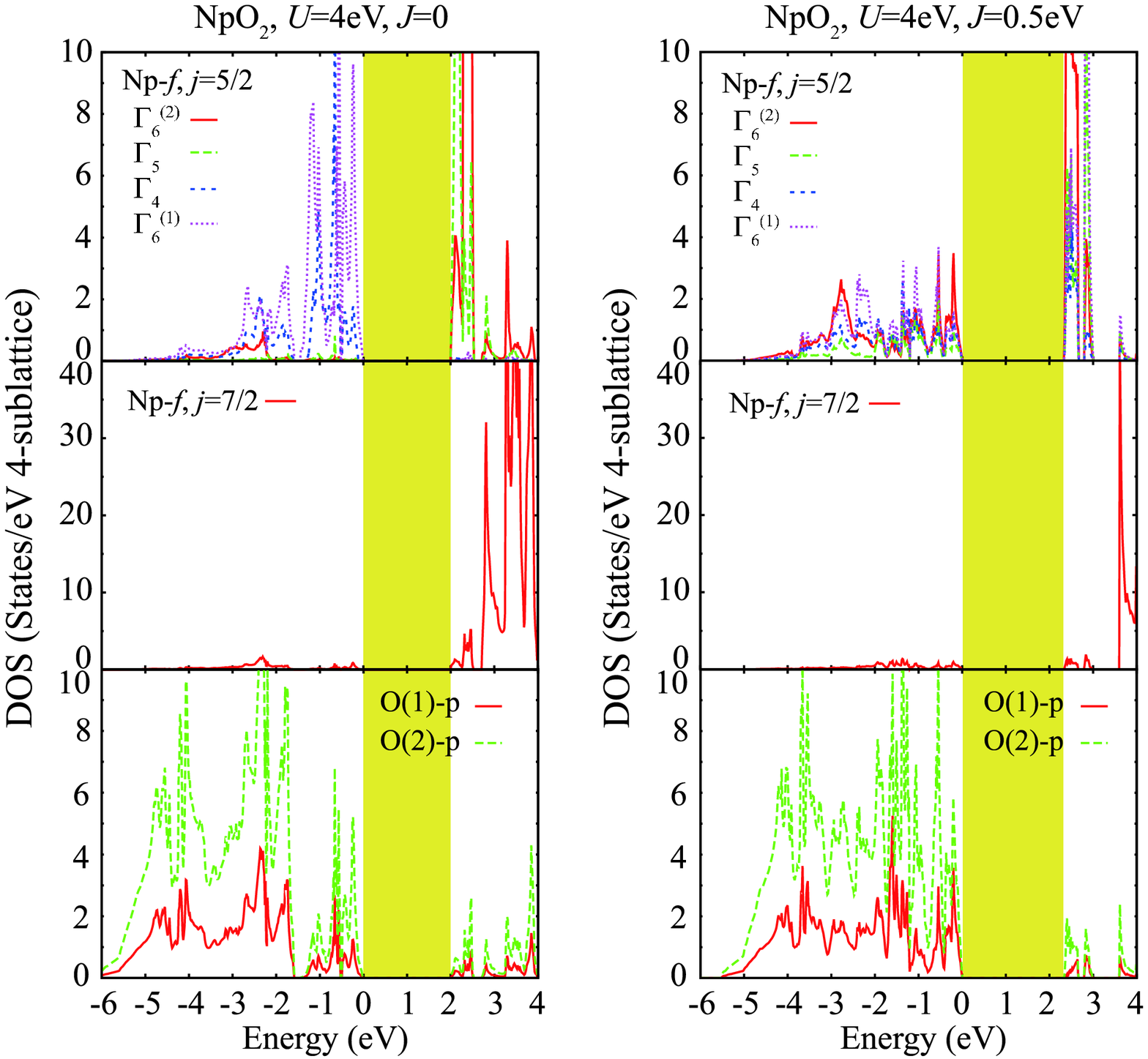}
  \caption{(Color online) The orbital-projected DOS components computed for the multipolar ordered phase of NpO$_2$, for $J=0$ (left) and $J=0.5$ eV (right).}
  \label{Fig:DOS_NpO2_Order}
 \end{center}
\end{figure}

\begin{figure}[tbh]
 \begin{center}
  \includegraphics[width=1.0\linewidth]{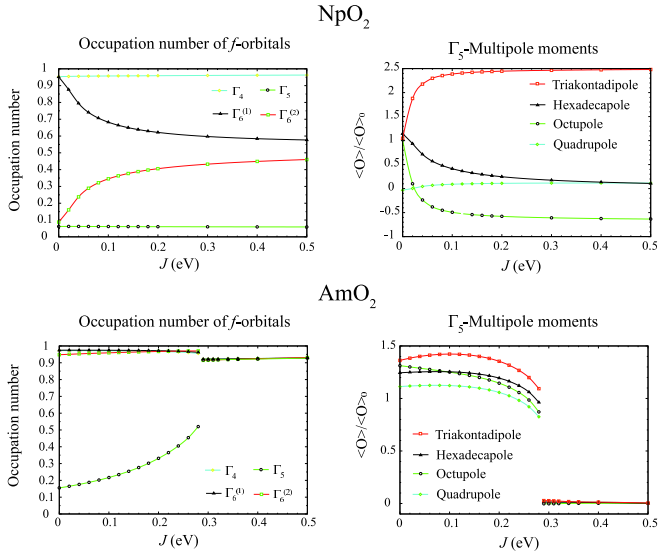}
  \caption{(Color online) 
  Left panels: calculated one-electron orbital occupation numbers as a function of exchange $J$ at $U=4$ eV, for the MPO states of NpO$_2$ (top) and AmO$_2$ (bottom).
Right panels: Calculated expectation values of the multipolar  (quadrupolar, octupolar, hexadecapolar, and triakontadipolar) order parameters in NpO$_2$ and AmO$_2$ as a function of $J$.}
\label{Fig:Multipole_NpAm}
 \end{center}
\end{figure}

\begin{figure}[tbh]
 \begin{center}
  \includegraphics[width=1.0\linewidth]{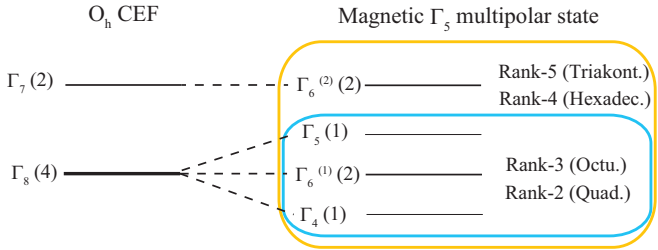}
  \caption{(Color online)
   Schematic picture showing the correspondence of the contributing $j=5/2$ orbitals and the active multipoles in the $\Gamma_5$ multipolar order.}
\label{Fig:CEF_G5MPO}
 \end{center}
\end{figure}

\begin{figure}[tbh]
 \begin{center}
  \includegraphics[width=0.7\linewidth]{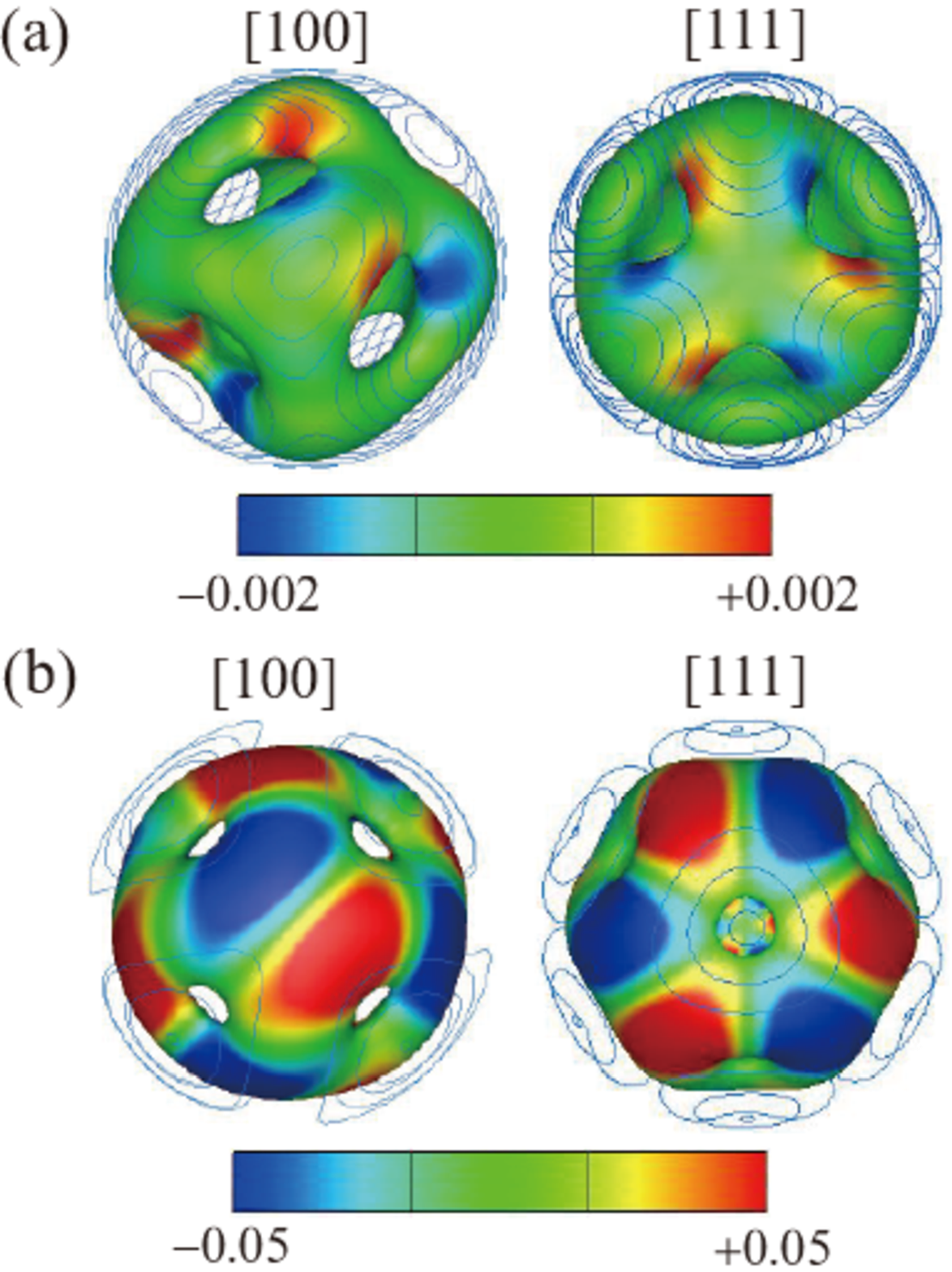}
  \caption{(Color online) Charge and magnetic distributions of the Np-5$f$ electrons in NpO$_2$, computed with (a) $U=4$ eV and $J=0$  and (b)  $U=4$ eV and $J=0.5$ eV. Details of the plots are as given in Fig.\ \ref{Fig:ChgSpn_U}.
}
\label{Fig:ChgSpn_Np}
 \end{center}
\end{figure}

 
Next, we consider NpO$_2$.
The obtained band structures and DOS are shown in Figs.\ \ref{Fig:BandDOS_NpO2_Order} and  \ref{Fig:DOS_NpO2_Order}, for $U=4$ eV and two $J$ values, 0 and 0.5 eV, respectively. For this actinide dioxide the computed electronic structure corresponds to the longitudinal 3$\boldsymbol{q}$ ordered state with $\Gamma_5$ local multipoles.

A schematic picture of how the one-electron $f$ orbital levels will split, depending on the  symmetry at the Np sites in NpO$_2$, has been provided in Fig.\ \ref{Fig:CEF}. 
The $\Gamma_{5}$ multipole order lowers the $O_h$ local symmetry to the $D_{3d}$ symmetry on the Np sites. Thus, the $| j$=$5/2, \Gamma_8 \rangle$ quartet splits into the two singlets $\Gamma_{4}$ and $\Gamma_{5}$ and one doublet $\Gamma_{6}^{(1)}$ in the MPO states. The two singlets $\Gamma_{4}$ and $\Gamma_{5}$ are degenerated under time-reversal invariance but not in the {\it magnetic} MPO states, which break the time-reversal symmetry. The two doublets derived from the $\Gamma_{7}$ and $\Gamma_{8}$ orbitals belong to the same symmetry and hybridize with each other in the MPO states.
A $\Gamma_{5}^{-}$ multipole is expected to be produced with the three $5f$ electrons occupying the $\Gamma_4$ singlet and the $\Gamma_6^{(1)}$ doublet, which are split from the $\Gamma_8$ quartet in the paramagnetic state. Accordingly, we thus choose, as initial density matrix, the one corresponding to the local $f$ states in which the $\Gamma_4$ singlet and the $\Gamma_6^{(1)}$ doublet are occupied by three $f$ electrons.  
In the longitudinal 3$\boldsymbol{q}$ structure, the induced $\Gamma_{5}$ multipoles obey a relation, for instance, $\langle O_{yz}\rangle$=$\langle O_{zx}\rangle$=$\langle O_{xy}\rangle$ for the Np (0,\,0,\,0) ion, $\langle O_{yz}\rangle$ =$-\langle O_{zx}\rangle$=$-\langle O_{xy}\rangle$ for the Np ion at (0,\,1/2,\,1/2), $-\langle O_{yz}\rangle$ =$\langle O_{zx}\rangle$=$-\langle O_{xy}\rangle$ for the Np (1/2,\,0,\,1/2) ion, and $-\langle O_{yz}\rangle$ =$-\langle O_{zx}\rangle$=$\langle O_{xy}\rangle$ for the Np (1/2,\,1/2,\,0) ion.
Reflecting the $D_{3d}$ local symmetry, the Np-$f$ orbital components in the DOS still keep the degeneracy for the $\Gamma_6$ doublets in the exotic magnetic multipole ordered states as shown in  Fig.\ \ref{Fig:DOS_NpO2_Order}. 

The top-right panel of Fig.\ \ref{Fig:Multipole_NpAm} shows the calculated $J$ dependence of the multipole moments $\langle O \rangle$ normalized 
to the multipole moments for the initial $f$ occupation, $\langle O \rangle_0$, as mentioned above.
The noncollinear multipole moments in the ordered state of NpO$_{2}$ are strongly affected by the value of the Hund's coupling $J$, which also affects the occupation difference of $f$-orbitals, see the top-left panel. At $J=0$, the $\Gamma_4$ singlet and the $\Gamma_{6}^{(1)}$ doublet are fully occupied. As $J$ increases, the $f$-electron on the $\Gamma_{6}^{(1)}$ doublet starts to transfer to the $\Gamma_6^{(2)}$ doublet through hybridization. This transfer strongly enhances the triakontadipole moment and suppresses the octupole moment in NpO$_{2}$. The occupation of the one-electron CEF orbitals and its relation to the multipolar order is schematically illustrated in Fig.\ \ref{Fig:CEF_G5MPO}.
Thus, the $\Gamma_5$-triakontadipole moment can be the leading order parameter in NpO$_2$.\cite{suzuki10_1}

The charge and magnetic distributions of the neptunium $5f$-wave function are plotted in Fig.\ \ref{Fig:ChgSpn_Np} for $U=4$ eV and two $J$ values, 0 and 0.5 eV. In contrast to the equivalent distributions of the U $5f$ wave function in Fig.\ \ref{Fig:ChgSpn_U}, the magnetic distribution corresponds to a vanishing atomic magnetic dipole moment along the [1\,1\,1] local axis after the space integration.  The calculated charge and magnetic distributions preserve the  $D_{3d}$ symmetry of NpO$_2$. The enhancement of the triakontadipole moment with increasing $J$ is reflected in the enhanced local magnetic moments on the charge isosurface in Fig.\ \ref{Fig:ChgSpn_Np}(b).

\begin{figure}[tbh]
 \begin{center}
  \includegraphics[width=0.88\linewidth]{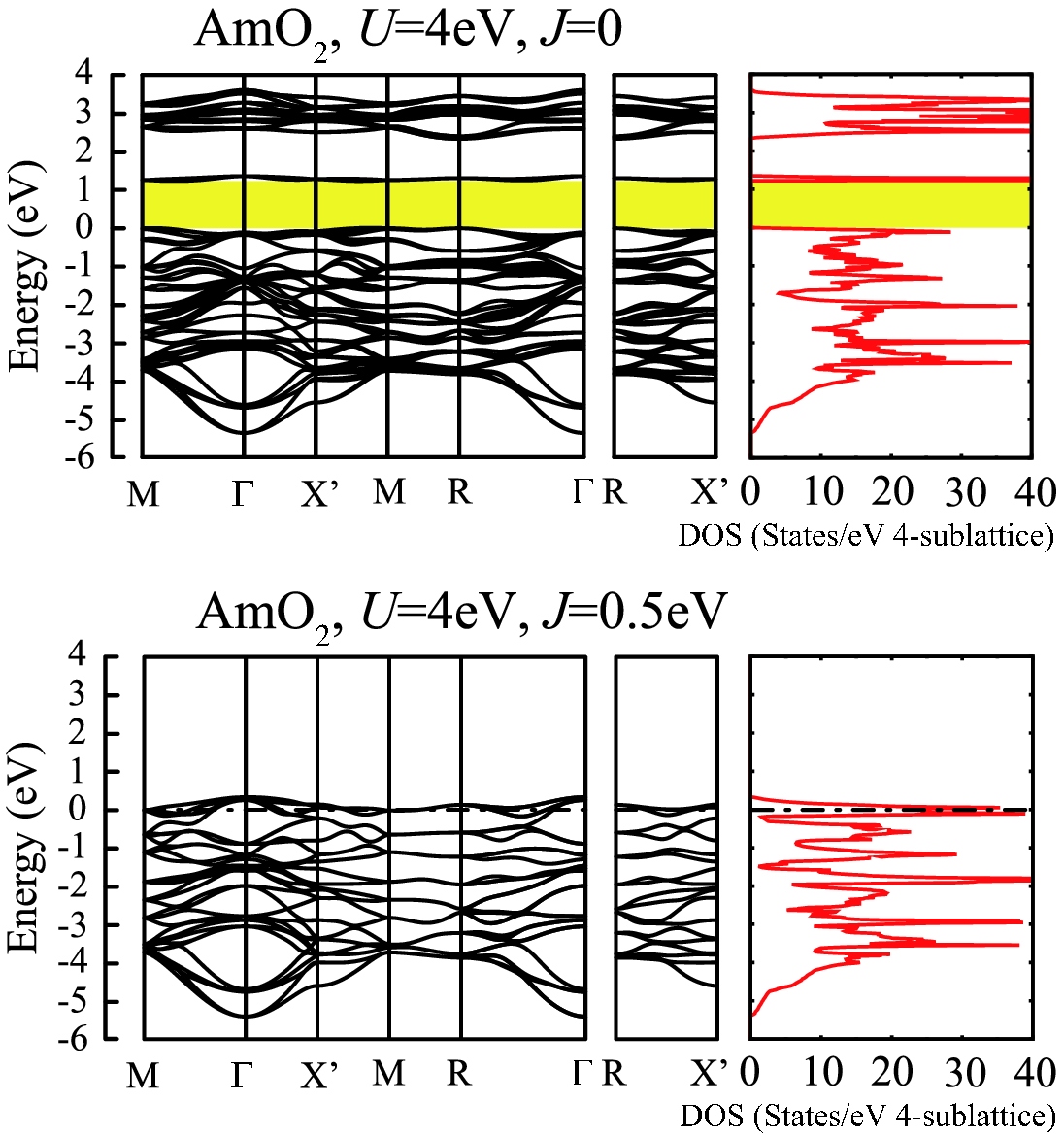}
  \caption{(Color online) Computed band structures and density of states in the longitudinal 3$\boldsymbol{q}$ $\Gamma_{5u}$ magnetic MPO state of AmO$_2$, calculated by the LDA+$U$ method ($U=4$ eV and $J=0$ or $J=0.5$ eV). }
\label{Fig:BandDOS_AmO2_Order}
 \end{center}
\end{figure}

\begin{figure}[tbh]
 \begin{center}
  \includegraphics[width=1.0\linewidth]{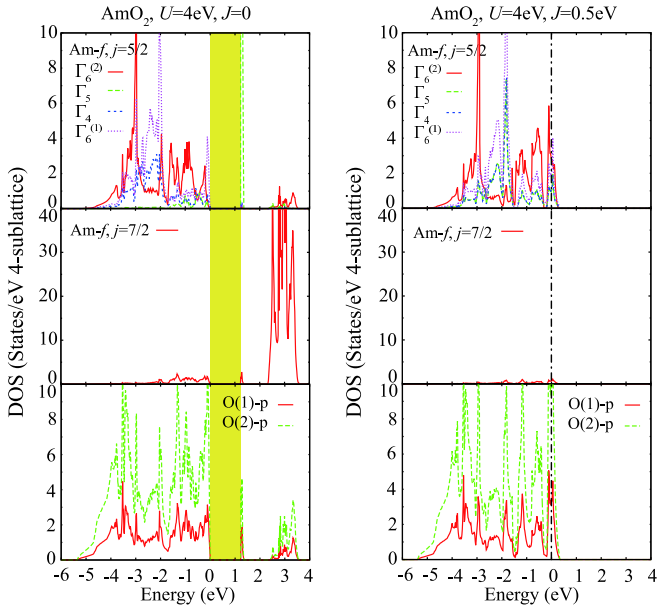}
  \caption{(Color online) As Fig.\ \ref{Fig:DOS_NpO2_Order}, but for the multipolar ordered phase of AmO$_2$.}
\label{Fig:DOS_AmO2_Order}
 \end{center}
\end{figure}

\begin{figure}[h]
 \begin{center}
 \includegraphics[width=0.7\linewidth]{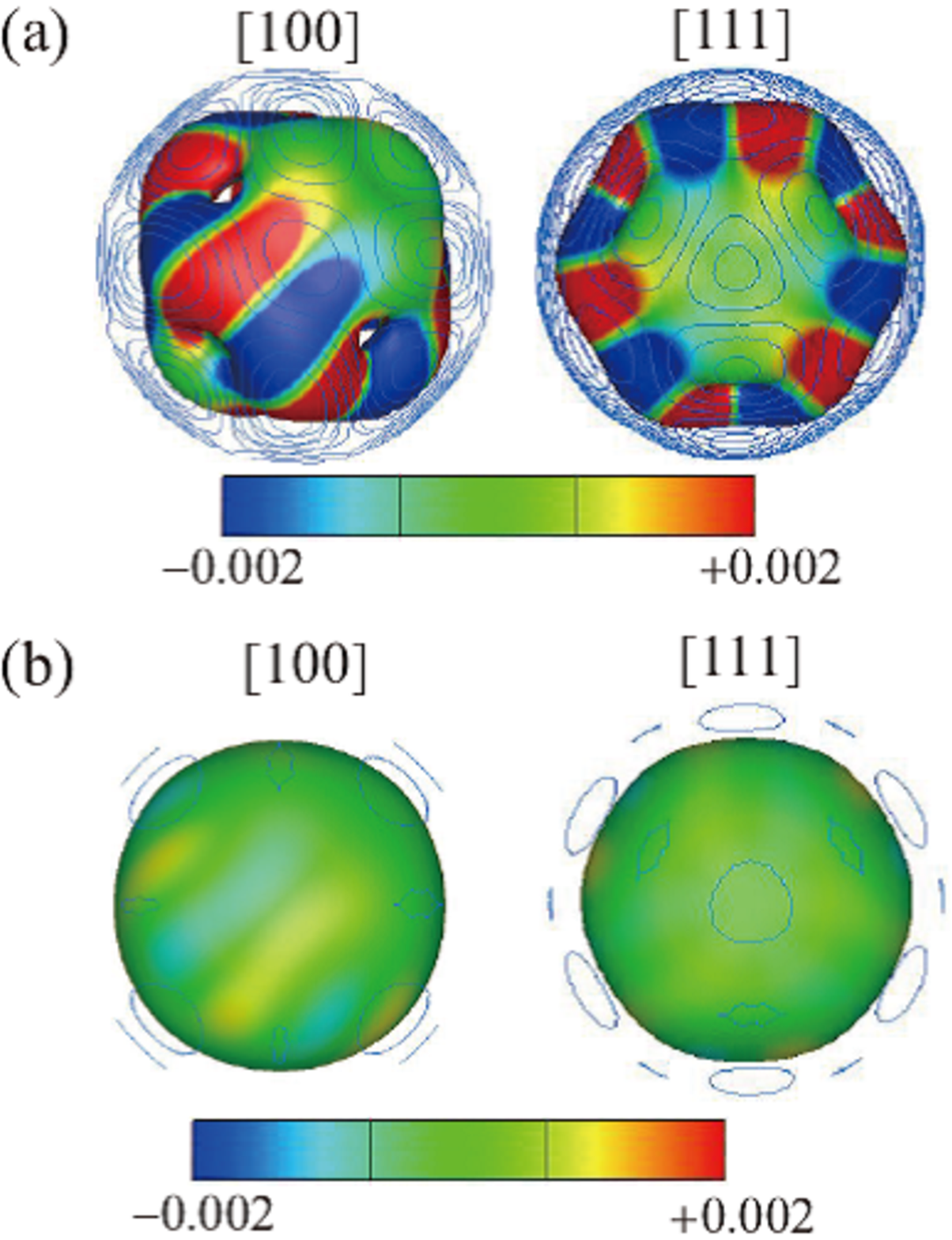}
  \caption{(Color online) Charge and magnetic distributions of the Am-5$f$ electrons in AmO$_2$, computed with (a) $U=4$ eV and $J=0$  and (b) by $U=4$ eV and $J=0.5$ eV. Details of the plots are as given in Fig.\ \ref{Fig:ChgSpn_U}.
}
\label{Fig:ChgSpn_Am}
 \end{center}
\end{figure}

Next, we consider AmO$_2$.
 The order parameter in the ground state of AmO$_2$ has not been determined experimentally yet. The recent theoretical study \cite{hotta09} of Hotta suggested the $\Gamma_8$ CEF ground state to explain the experimental facts.\cite{tokunaga10} It is hence plausible to have order parameters which are similar to those of NpO$_2$. 
 We thus choose initial density matrixes, so as to have finite $\Gamma_{5}$ multipole moments, corresponding to the local $5f$ state which has filled $\Gamma_6^{(1)}$ and $\Gamma_6^{(2)}$ doublets as well as the $\Gamma_4$ singlet in the calculations for ordered states in AmO$_2$.
 Then, our calculations lead to the solution of the longitudinal 3$\boldsymbol{q}$ ordered state with  $\Gamma_5$-local multipoles only for small $J$ as shown in Figs.\ \ref{Fig:Multipole_NpAm}.  
The obtained band structures and DOS are shown in Fig.\ \ref{Fig:BandDOS_AmO2_Order} and  \ref{Fig:DOS_AmO2_Order}, for $U=4$ eV and again two $J$ values, 0 and 0.5 eV, respectively. 
The charge and magnetic distributions of the americium $5f$-wave function are plotted in Fig.\ \ref{Fig:ChgSpn_Am} for $U=4$ eV and two $J$ values, 0 and 0.5 eV. These are considerably different from the equivalent distributions computed for the Np $5f$ wave function in Fig.\ \ref{Fig:ChgSpn_Np}, reflecting the substantially different constitution of the high-rank multipoles in AmO$_2$ from that of NpO$_2$, see Fig.\ \ref{Fig:Multipole_NpAm}.

We have found that the multipolar ordering is responsible for stabilizing the insulating ground state for small values of $J$ in AmO$_2$, like in NpO$_2$; the resulting gap is indicated by the shaded area in Fig.\ \ref{Fig:BandDOS_AmO2_Order}.
Meanwhile, the constitution of the high-rank multipoles in AmO$_2$ differs substantially from the one of NpO$_2$, see Fig.\ \ref{Fig:Multipole_NpAm}.
 The $\Gamma_5$-multipoles in AmO$_2$ are rather insensitive for small $J$ values (bottom-right panel), whereas they decrease for $J$ values larger than 0.1 eV and show an abrupt disappearance of the multipole order for $J$ values slightly smaller than 0.3 eV. This striking difference of the $J$ dependence of the multipoles in AmO$_2$ from those of NpO$_2$ stems from the fact that the $| j\!=\!5/2, \Gamma_7 \rangle$ orbitals are completely occupied in the $f^5$ state of the Am$^{4+}$ ion and there is no state available to couple to it through the Hund's coupling $J$.
 Furthermore, the calculated energy gap for AmO$_2$ is found to depend strongly on the Hund's coupling $J$, see Fig.\  \ref{Fig:BandDOS_AmO2_Order}. Clearly, with increasing $J$ a magnetically ordered solution becomes more favorable. 
 A similar sensitivity was not observed for UO$_2$ nor for NpO$_2$. As can be seen from Fig.\ \ref{Fig:BandDOS_AmO2_Order}, the energy gap in AmO$_2$ decreases with increasing $J$ and disappears simultaneously with the disappearance of the multipole order as discussed above. This implies that a large Hund's rule $J$ can make the insulator solution as well as the longitudinal 3$\boldsymbol{q}$ ordered state of $\Gamma_{5}^{-}$ multipoles unstable in AmO$_2$. 



\section{\label{sec:level4} Conclusions}

We have investigated the origin of the gap formation in the actinide dioxides. 
The origin of the insulating gap formation is found to lie in the strong on-site Coulomb repulsion of the $5f$ electrons and strong spin-orbit interaction in the relevant $f$ orbitals in the AnO$_2$ compounds.
 LDA+$U$ calculations for a non-long-range ordered state reproduce well the energy gaps following the singlet CEF ground states in PuO$_2$ and CmO$_2$.
On the other hand, the insulating ground states in UO$_2$, NpO$_2$, and AmO$_2$ are obtained only by allowing for the symmetry lowering that can give rise to the ordered states. 
Thus, the strong correlation is necessary to describe the anisotropic $f$ ground states in AnO$_2$.
The energy gaps and magnetic properties are correctly reproduced within the accepted range of the parameters, $U$ and $J$, by taking the proper magnetic space symmetries in the calculations.
Especially,  using  values of $J$ in the acceptable range produces magnetic moments which are reduced from the one expected from the CEF ground states,  yet with the appropriate experimental energy gap in UO$_2$. In addition the contribution from the electric quadrupole moments is enhanced with increasing $J$.
We also showed that the active multipoles in these ordered states are closely related to the orbital occupation, and the higher rank $\Gamma_{5}$ multipole ordered states in AmO$_2$ have quite different constitution of the multipoles from NpO$_2$. 
 Whereas Hund's coupling $J$ enhances the energy gap in NpO$_2$ together with changing the constitution of high-rank multipoles, it makes the insulator states with 3$\boldsymbol{q}$ longitudinal $\Gamma_{5}$ multipole order unstable in AmO$_2$.
Further experimental investigations are required and encouraged to verify the here-computed ground state properties  of AmO$_2$ and CmO$_2$.

\acknowledgments
{We thank R.\ Caciuffo, S.\ Kambe, and Y.\ Tokunaga for valuable discussions.
This work has been supported by JSPS KAKENHI Grant Number 23246174 and 24540369, the Swedish Research Council (VR), the Joint Research Center of the European Commission, Svensk K{\"a}rnbr{\"a}nslehantering AB (SKB), Swedish National Infrastructure for Computing (SNIC), the Supercomputer Center of the Institute for Solid State Physics at the University of Tokyo and by the Japan Atomic Energy Agency (JAEA). Part of this work was performed at LBNL under the LDRD program, which is supported by the Director, Office of Science, of the U.S. Department of Energy under Contract No. DE-AC02-05CH11231.}


\begin{thebibliography}{99}
\bibitem{note}
This is the conventional labeling of multipolar symmetry, which should not be confused with the labeling of the CEF one-electron orbitals.
\end{thebibliography}
\end{document}